# High-temperature structural phase transition in multiferroic $LiCu_2O_2$

S. A. Ivanov[a,b], A. A. Bush[c], K. E. Kamentsev[c], E. A. Tishchenko[d],
M. Ottosson[e], R. Mathieu[b], P. Nordblad[b]

a - Department of Inorganic Materials, Karpov' Institute of Physical Chemistry, RU-105064, Moscow, Russia
b - Department of Engineering Sciences, Uppsala University, Box 534, SE-751 21 Uppsala, Sweden
c - Moscow State Technical University Radioengineering, Electronics and Automation, RU-119434, Moscow, Russia
d - Kapitza Institute for Physical Problems RAS, RU-119334, Moscow, Russia
e - Department of Chemistry, Uppsala University, Box 538, SE-751 21 Uppsala, Sweden

$LiCu_2O_2$ single crystals were studied in the temperature range 300 – 1100 K by means of heating–cooling curves of differential thermal analysis (DTA), thermogravimetry (TG), X-ray powder diffraction and electrical measurements. A reversible first-order phase transition between orthorhombic and tetragonal phases was found to take place at 993 K. At the transition, a peak is observed in the DTA curves, as well as jumps of the unit cell parameters and electrical resistivity. Considering the crystal structure of $LiCu_2O_2$ and the entropy change associated with the phase transition, it is concluded that the phase transition is related to processes of order-disorder of the $Cu^{2+}$ and $Li^+$ cations onto their crystallographic positions.

## I. INTRODUCTION

The frustrated magnet $LiCu_2O_2$ [1 - 6] is one of the few systems among cuprates in which magnetism and ferroelectricity coexist and electric polarization can be reversibly flipped with an applied magnetic field. $LiCu_2O_2$ contains equal amounts of nonmagnetic $Cu^+$ and magnetic $Cu^{2+}$ (S=1/2) cations which are fully ordered onto their structural sites at room temperature [7, 8], where the compound has orthorhombic crystal structure (space group *Pmna*, $a= 5.730(1)$, $b = 2.8606(4)$, $c = 12.417(2)$, z = 4, see Fig. 1).

$LiCu_2O_2$ is a model system for investigations of low-dimensional magnetism [9 – 18], owning to the double copper - oxygen chains along the *b* axis forming a zigzag ladder-like structure. The ladders are isolated from each other by both $Li^+$ ions and layers of nonmagnetic $Cu^+$ [7 - 18] (Fig. 1). Exchange interactions between $Cu^{2+}$ in these chains cause two consecutive magnetic phase transitions at $T_{N1} = 24.6$ K and $T_{N2} = 23.2$ K with formation of a collinear modulated spin structure below $T_{N1}$ followed by a non-collinear helicoidal modulated spin structure below $T_{N2}$ [2, 4 - 6, 15, 18]. A strong magnetic - ferroelectric coupling has been evidenced below $T_{N2}$ [11-13]. A spontaneous electric polarization $P_s$ is induced along the *c*-axis of the crystals by the non-collinear helicoidal modulated spin structure stabilized below 23.2 K. The direction of $P_s$ can be changed by application of an external magnetic field [11-13].

Furthermore, LiCu$_2$O$_2$ crystals also exhibit threshold electrical switching from a high to a low resistance state and characteristic S-shaped I-V curves containing a region of negative differential resistivity [19, 20]. The rather low values of the critical voltage of switching from the high to the low resistance state suggest their potential use as active elements of switching devices, controllable inductive elements, and relaxational generators. Interestingly, the crystal chemistry of LiCu$_2$O$_2$ and high temperature Cu-based superconductors are also quite similar [21, 22].

Most of the earlier investigations of LiCu$_2$O$_2$ have been carried out at low temperatures (T < 300 K) studying the interplay of spin, charge, and lattice degrees of freedom of a model multiferroic system. Also, complete data about the high-temperature region of the phase diagram of the Li$_2$O – CuO – Cu$_2$O system, and related physical properties of LiCu$_2$O$_2$ are lacking in the literature as no systematic studies have been carried out above room temperature. Here we report on high-temperature thermophysical and X-ray powder diffraction studies as well as measurements of electrical resistivity on single crystal samples of LiCu$_2$O$_2$. A reversible first-order transition between orthorhombic and tetragonal phases was observed at 993K, tentatively associated with cation order-disorder phenomena.

## II. EXPERIMENTAL

The single-crystals of LiCu$_2$O$_2$ used in the present experiments were grown by the flux crystallization method by slow cooling of the melt 80CuO·20Li$_2$CO$_3$ in alumina crucibles. The growth parameters have been optimized and the corresponding details were described earlier [20]. The crystals have black color and they are characterized by lamellar habitus with developed faces {001} (along these faces crystals showed a perfect cleavage) and less developed faces {210}. The size of the crystals reached 4x10x10 mm.

X-ray powder diffraction (XRPD) data at room temperature were collected on a Bruker D8 Advance diffractometer (Ge monochromatized Cu K$_{\alpha1}$ radiation, Bragg-Brentano geometry, DIFFRACT plus software) in the 2θ range 10 – 140° with a step size of 0.02° (counting time was 15 s per step). The slit system was selected to ensure that the X-ray beam was completely within the sample for all 2θ angles.

Phase identification and purity of the powder sample was checked from XRPD patterns at room temperature. Powder diffraction pattern of received crystals showed the expected spectra without any impurity lines (Fig. 2). All reflections in the room temperature XRPD data were compatible with a single-phase orthorhombic structure (space group *Pnma*) with *a* = 5.7295(4), *b* = 2.8599(3), *c* = 12.4161(5) Å. These lattice parameters for LiCu$_2$O$_2$ at room temperature are in a reasonable agreement with those published in refs. [7, 8] and ICDD database (PDF card 80-2301). The slight disagreement with published cell parameters of LiCu$_2$O$_2$ could be explained by the different sample preparation methods.

Thermogravimetric studies were performed using Q - 1500D TG/DTA (thermogravimetry/differential thermal analysis) Paulic – Erdei system. The powder sample was placed in a Pt crucible and α-alumina Al$_2$O$_3$ was used as thermally inert material (the reference sample). The temperature change was linear, 10 K/min. For thermogravimetric and x-ray diffraction studies selected single crystals of LiCu$_2$O$_2$ were crushed into pow-der in an agate mortar. High-temperature x-ray diffraction studies were performed on a Si substrate, which was covered with several drops of the resulting powder suspension of $LiCu_2O_2$ in ethanol, leaving randomly oriented crystallites after drying. In this case possible parasitic phases can be leached in 10% diluted $HNO_3$. The X-ray diffraction measurements at variable temperature in the temperature range 298 – 1073 K were carried out on the powdered sample using an Anton Paar XRK 900 chamber. Helium was used as protective gas during high-temperature experiments, as the decomposition of the phase $LiCu_2O_2$ take place under heating in air above 620 K [20].

The magnetic properties of the crystals were measured using a SQUID magnetometer from Quantum Design Inc. The magnetization M was recorded in a small magnetic field (H = 20 Oe) as a function of temperature (5-300 K) in constant field using zero-field-cooled and field-cooled protocols. The temperature dependence of magnetization shown in Fig. 3 is similar to earlier data obtained on $LiCu_2O_2$ [4, 6, 9, 10, 12, 13, 15 - 18]. The M(T) curves measured in zero-field- and field-cooled conditions sharply decrease below 50 K, and (antiferro)magnetic transitions can be observed in the temperature-derivative of the magnetization (see Ref. [18]). The divergence between zero-field and field cooled curves below 150 K may reflect the onset of weak spontaneous magnetization reported in Ref. [17].

Measurements of electrical resistivity were carried using an LCR/ESR meter MT4090 (the Motech Industries) in the temperature range of 300 – 1100 K on constant (DC) and on variable (AC) current with frequencies 0.1, 1, 10, 100 and 200 kHz. For the electro-physical studies, the plates were cut out from crystals with the basic planes parallel to the faces (001) and (210). The area of basic face and thickness of the plates were equal to ~10 mm$^2$ and 1-2 mm respectively. Ag-electrodes were attached on the basic surfaces of the plates.

## III. RESULTS AND DISCUSSION

### A. Thermogravimetric analysis

Under heating of $LiCu_2O_2$ powder in air atmosphere a two-stage thermal decomposition of the $LiCu_2O_2$ phase in CuO and $Li_2CuO_2$ binary mixture (Fig. 4a) occurs in the range 533 – 773 K. This result is in accordance with earlier findings in Ref. [20]. The increase in weight of the sample at decomposition corresponds to transformation of all $Cu^+$ to $Cu^{2+}$ state in the studied phase $LiCu^{2+}Cu^+O_2$. Thus, the thermogravimetric analysis confirms stoichiometry of $LiCu_2O_2$ phase in the studied crystals. At higher temperatures, in the range of thermal stability of the phase $LiCu_2O_2$ (T > 1058 K), a reduction of weight corresponding to reverse reaction of the $LiCu_2O_2$ phase formation from the products of its decomposition is observed.

During heating of $LiCu_2O_2$ powder in argon atmosphere the phase preserves its thermal stability up to the melting temperature (~1300 K). However, during heating the DTA curves show a very clear endothermic signal at 993 K, evidencing a first order transition. On cooling, a corresponding exothermic signal is observed at 983 K, indicating a reversible phase transition (see Fig. 4b). The observed peaks are well reproduced in repeated measurements (Fig. 4b). Subsequent powder diffraction experiments show that the DTA peak can be attributed to a transformation into a tetragonal structure.

## B. High temperature X-ray powder diffraction

Variable temperature X-ray diffraction patterns recorded between 298 K and 1073 K show a structural change around 983 – 993 K ($T_{PT}$). We show portions of the diffractograms of room temperature and high temperature phases in Fig. 2. It was found that important changes in the intensity of some reflections take place above $T_{PT}$. Several corresponding (hkl) reflections with h=2n+1, identified for the low temperature orthorhombic polymorph, disappeared confirming the structural phase transition to a tetragonal phase. The structural phase transition is reversible (not associated with a compositional change) and is of first order. The temperature of this phase transition is consistent with the results of thermal analysis. Closer inspection of obtained diffraction data revealed that the new unit cell is tetragonal with the same $c$ and $a$ axes, with a cell parameter $a$ that is one half of the former one. The unit cell dimensions were determined from the high-temperature set of the observed reflections and could be indexed on the base of the tetragonal structure with $a$ = 2.8959(4) and $c$ = 12.5490(6) Å (1053 K). The extinction conditions indicated that a primitive space group of symmetry is the most probable for high temperature tetragonal form of $LiCu_2O_2$.

Figure 5 illustrates the thermal evolution of lattice parameters for $LiCu_2O_2$ in the high temperature range. As the temperature is raised up to $T_{PT}$, the lattice parameters increase gradually, almost linearly with thermal expansion coefficients along *a*, *b* and *c* directions: $\alpha_a$= 11.2x10$^{-5}$, $\alpha_b$= 4.2x10$^{-5}$ and $\alpha_c$= 19.2x10$^{-5}$ 1/K. The thermal expansion is found to be rather anisotropic, strongest along *c*-direction and weakest along *b*-direction. The unit cell parameters undergo an important change around $T_{PT}$ showing its discontinuous (step-wise) increase. Above $T_{PT}$ the *a* cell parameter becomes equal to parameter *b*. The order parameter for the phase transition can be defined as "orthorhombic strain" **(*a*−2*b*)/(*a*+2*b*)**.

## C. Electrophysical measurements

During heating a smooth decrease of the resistivity is observed, until the temperature reaches the phase transition temperature $T_{PT}$ = 993 K in the vicinity of which a sharp reduction of resistivity takes place. Such sharp step-like changes of resistivity are observed in measurements both along and perpendicular to the crystallographic *c*-axis of the crystal in heating and cooling modes (Fig. 6). It can be that the resistivity weakly depends on frequency, and DC and AC resistivity have close values. Considering Ag electrodes as blocking for an ionic component of conductivity, it is possible to conclude that prevailing contribution has an electronic origin.

In the range of 873 – 993 K hysteresis is observed in the resistivity: the values of ρ(T) on heating lie appreciably below those on cooling; below ~873 K no hysteresis is observed. The temperature dependence of the resistivity is linear in ln ρ vs. 1/T plots (ln here and below refers to as the natural logarithm) in the range of 300 - 550 K, yielding activation energies, $E_a$ = 0.07 eV and 0.17 eV along and perpendicular to *c* - axis, respectively.

### D. Analysis of the phase transition

The results of the thermogravimetric, x-ray diffraction and electrical resistivity measurements indicate that $LiCu_2O_2$ crystals undergo a reversible first-order phase transition with symmetry change from orthorhombic to tetragonal at 993 K. The orthorhombic symmetry of the low-temperature phase is mainly related to the $Cu^{2+}$ - O and $Li^+$ - O chains extended along the axis *b*. Increase of symmetry of a phase to tetragonal is possible only at the statistical or ordered redistribution of cations of $Cu^{2+}$ and $Li^+$ on their structural positions leading to averaging of specified chains in the directions along the *a* and *b* axes (see Fig. 1).

The area, $\Delta A$, under the DTA peak at the phase transition (Fig. 4b) allows an estimate of the enthalpy (heat) of transition $\Delta H$ : $\Delta H = K \cdot M \cdot \Delta A$, where $K$ is a calibration factor and $M$ the mole mass. For calibration of the area we used the thermograms of $BaCO_3$ powder which was recorded in similar conditions and tabulated data [23] on the heat of the phase transition in $BaCO_3$ at 1079 K (16.2 kJ/mol). The value of the associated entropy change $\Delta S$ could be calculated as $\Delta S = \Delta H/T_{PT} = 4.1$ J/(mol·K) = R ln(1.7). The obtained value of $\Delta S$ indicates a considerable orientational disorder in the high-temperature phase [24] and therefore we can conclude that the observed phase transition belongs to order-disorder type. Judging from shape of curve DTA (Fig. 4b) a possible order - disorder process in the $Li^+$ - $Cu^{2+}$ sublattice begins at about 870 K, the further temperature increasing accompanied by a gradual increase of the disorder which completes drastically at the temperature of the phase transition.

The possibility of interchangeability of $Li^+$ and $Cu^{2+}$ cations in $LiCu_2O_2$ stems from the similarity of their ionic radii of $Li^+$ (0.73 Å) and $Cu^{2+}$ (0.71 Å) [25] and the same $MO_5$ cation - oxygen polyhedral environment for both cations. Such disorder of $Li^+$ and $Cu^{2+}$ cations in structural positions has earlier been established by X-ray structural analysis at room temperature in the crystals $LiCu_3O_3$ crystals whose chemical formula can be represented as $(Li^+_{0.2}Cu^{2+}_{0.8})Cu^+(Li^+_{0.8}Cu^{2+}_{1.2})O_3$ [26]. Disordering of the cations in the $Cu^{2+}$ and $Li^+$ sublattices during the phase transition is responsible for the decrease of the resistance. The amount of cation order-disorder is different on heating and cooling, yielding the observed hysteresis in the resistivity above 873 K.

Polysynthetic twinning observed in $LiCu_2O_2$ single crystals [1 – 4, 7, 8, 12 – 14, 16, 17] is connected with the lowering of the symmetry from tetragonal to orthorhombic through the here reported transition. This suggests ferroelastic properties below 993 K [27], and hence ferroic order at temperatures well above those at which magnetism induces electric polarization.

## IV. CONCLUSIONS

In summary, our high-temperature studies of $LiCu_2O_2$ single crystals reveal a reversible first-order phase transition around $T_{PT} = 993$ K. In the vicinity of $T_{PT}$ well-defined peaks on DTA curve, an abrupt change of lattice parameters with symmetry change from orthorhombic to tetragonal and stepwise changes of the electrical resistivity were observed. The orthorhombic room temperature modification of $LiCu_2O_2$ crystals can be transformed to the tetragonal structure by heating and this phase transition is found to be reversible.

The specifics of the LiCu$_2$O$_2$ crystal structure and the value of entropy change at the transition suggest that the processes of order - disorder of Li$^+$ and Cu$^{2+}$ cations in their structural positions are responsible for the phase transition. The final distribution of the specified cations between possible structural positions depends on the sample preparation conditions including the cooling rate from high temperatures. This distribution also significantly influences the electrical characteristics of the crystals at room temperature.

**ACKNOWLEDGMENTS**

Financial support from the Russian Foundation for Basic Research and the Swedish Research Council is gratefully acknowledged. The authors are grateful to R. Berger for numerous insightful discussions on the subject and for his constructive comments and to M. Hudl and P. Anil Kumar for technical support.

# REFERENCES

1. R. Berger, A. Meetsma, and S. van Smaalen, J. Less-Common Metals, **175**, 119 (1991).
2. R. Berger, P. Önnerud, and R. Tellgren, J. Alloy Compds, **184**, 315 (1992).
3. A.M. Vorotynov, A.I. Pankrats, G.A. Petrakovskii, K.A. Sablina, W. Paszkowicz, H. Szymczak, J. Exp. Theo. Phys. **86**, 1020 (1998).
4. F.C. Fritschij, H.B. Brom, R. Berger. Sol. State. Commun. **107**, 719(1998).
5. B. Roessli, U. Staub, A. Amato, D. Herlach, P. Pattison, K. Sablina, G.A. Petrakovskii, Physica B **296**, 306 (2001).
6. S. Zvyagin, G. Cao, Y. Xin, S. McCall, T. Caldwell, W. Mouton, L.-C. Brunel, A. Angerhofer, and J.E. Crow, Phys. Rev. B **66**, 064424 (2002)
7. T. Masuda, A. Zheludev, A.Bush, M. Markina, and A. Vasiliev, Phys. Rev. Lett. **92**, 177201 (2004).
8. T. Masuda, A. Zheludev, B. Roessli, A. Bush, M. Markina, and V. Vasiliev, Phys. Rev. B **72**, 014405 (2005).
9. P. Zheng, J.-L. Luo, D. Wu, S.-K. Su, G.-T. Liu, Y.-C. Ma, and Z.-J. Chen, Chin. Phys. Lett. **25**, 3406 (2008).
10. L.E. Svistov, L.A. Prozorova, A.M. Farutin, A.A. Gippius, K.S. Okhotnikov, A.A. Bush, K.E. Kamenzev, E.A. Tischenko, J. Exp. Theo. Phys. **108**, 1000 (2009).
11. S. Park, Y.J. Choi, C.L. Zhang, S.-W. Cheong. Phys. Rev. Lett. **98**, 057601 (2007).
12. A. Rusydi, I. Mahns, S. Müller, M. Rübhausen, S. Park, Y.J. Choi, C.L. Zhang, S.-W. Cheong, S. Smadici, P. Abbamonte, M.v. Zimmermann, G.A. Sawtzky, Appl. Phys. Lett. **92**, 262506 (2008).
13. S. Seki, Y. Yamasaki, M.Soda, M. Matsuura, and Y. Tokura, Phys. Rev. Lett. **100**, 127201 (2008).
14. Y. Yasui, K. Sato, Y. Kobayashi, and M. Sato, J. Phys. Soc. Jpn. **78**, 084720 (2009).
15. M. Kobayashi, K. Sato, Y. Yasui, T. Moyoshi, M. Sato, and K. Kakurai, J. Phys. Soc. Jpn. **78**, 084721 (2009).
16. H. C. Hsu, W. L. Lee, J.-Y. Lin, H. L. Liu, and F. C. Chou, Phys. Rev. B **81**, 212407 (2010).
17. E. A. Tishchenko, O. E. Omelyanovskii, A. V. Sadakov, D. G. Eshchenko, A. A. Bush, and K. E. Kamenzev, Solid State Phenomena, Trend in magnetism Vols. **168-169**, p. 497-500. (2011).
18. A. A. Bush, V. N. Glazkov, M. Hagiwara, T. Kashiwagi, S. Kimura, K. Omura, L. A. Prozorova, L. E. Svistov, A. M. Vasiliev, and A. Zheludev, Phys. Rev. B **85**, 054421 (2012).
19. A. A. Bush, K. E. Kamentsev. Phys. Solid State **46**, 445 (2004).
20. A. A. Bush, K. Kamentsev, E. A.Tishchenko Inorg. Mater. **40**, 44 (2004).
21. Crystal chemistry of high-$T_c$ superconducting copper oxides. By B. Raveau, C. Michel, M. Hervieu, and D. Groult, Springer, Berlin 1991, X, 331 pp., hardcover, DM 149, ISBN 3-540-51543.
22. M. Uehara, T. Nagata, J. Akimitsu, H. Takahashi, N. Môri, K. Kinoshita. J. Phys. Soc. Jpn. **65**, 2764 (1996).
23. Landolt-Börnstein. Zahlenwerde und Funktionen aus Physik, Chemie, Geophysik, Astronomie, Technik. 4 Teil. Berlin. Springer-Verlag. 1961.

24. K. Binder, Rep. Prog. Phys. **50**, 783 (1987).
25. R. D. Shannon. Acta Cryst. **A32**, 731 (1976)
26. S. J. Hibble, K. Kohler, A. Simon, and S. Paider. J. Solid State Chem. **88**, 534 (1990).
27. K. Aizu, Phys. Rev. B **2**, 754-772 (1970).

**Figure captions**

Figure 1. Orthorhombic crystal structure of the low-temperature form of $LiCu_2O_2$ (space group *Pnma* - №62, *a*=5.73 Å, *b*=2.86 Å, *c*=12.41 Å, z=4).

Figure 2. X-ray diffractorgams at room temperature, 873 K and 1053 K (the reflections of the low temperature form which are absent in the diffractogram of the high temperature form are indicated by a *).

Figure 3. Temperature dependencies of the magnetization of crystal $LiCu_2O_2$ measured in conditions of field-cooled (FC) and zero-field-cooled (ZFC) in a magnetic field of 20 Oe.

Figure 4. Thermograms of powder of a $LiCu_2O_2$ single crystal, measured in air - a) and in argon – b) atmosphere (DTA – the curve of differential thermal analysis, m – mass of the sample, T – temperature, t – time). Relevant temperatures are indicated on the graphs.

Figure 5. Temperature dependence of the unit cell parameters of $LiCu_2O_2$.
The vertical error bars represent the uncertainties in the XRPD results.

Figure 6. Temperature dependencies of resistivity, $\rho$, of $LiCu_2O_2$ crystals measured along (a, c) and perpendicular (b, c) to the c –axis during heating and cooling of the crystals. In panels (a) and (b), the curves labelled 1 correspond to dc-resistivity and the coinciding curves labelled 2-6 to frequencies: 0.1, 1, 10, 100 and 200 kHz, respectively.

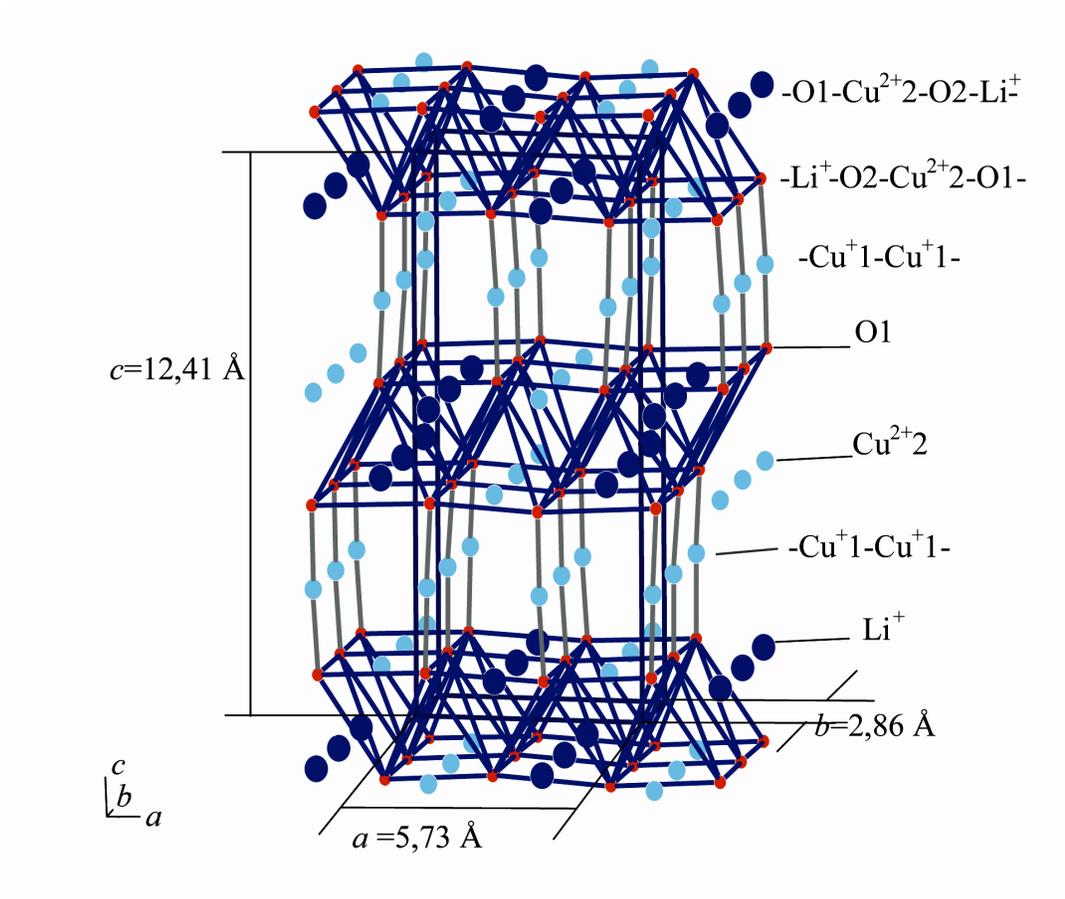

**Figure 1**

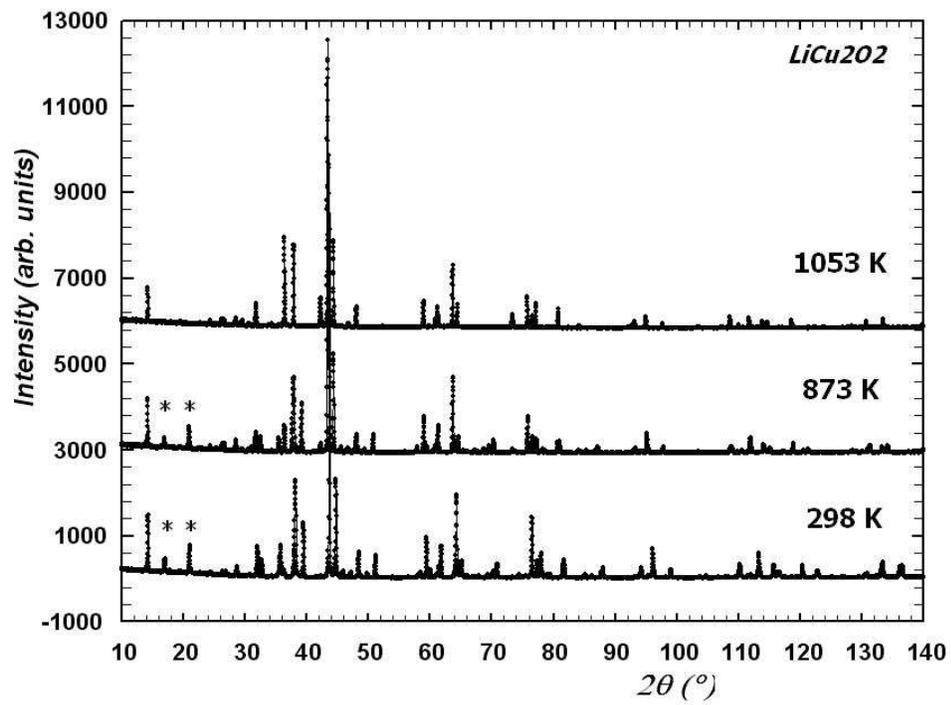

**Figure 2**

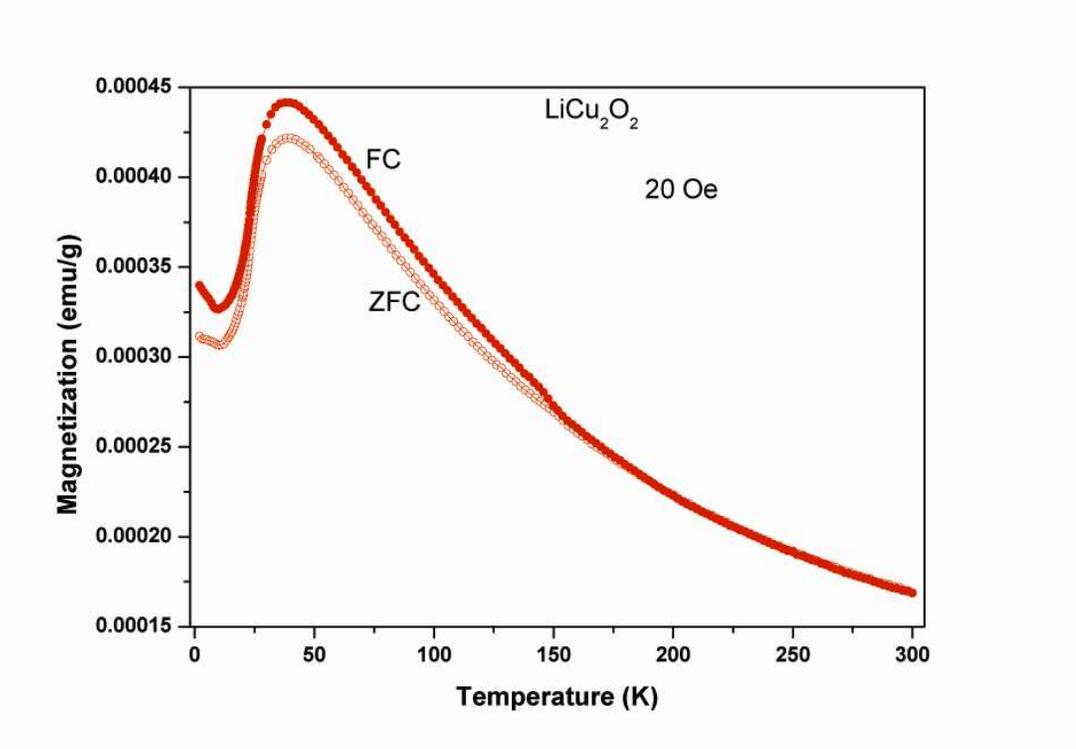

**Figure 3**

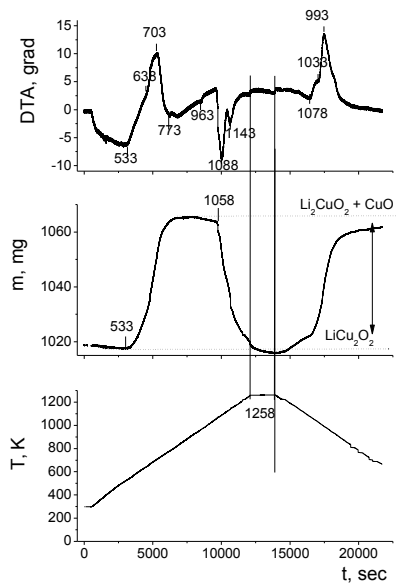
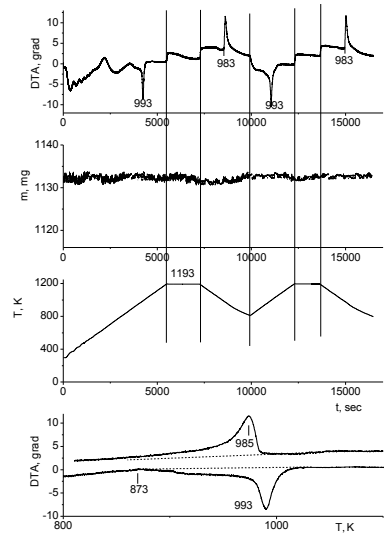

a)          b)

**Figure 4**

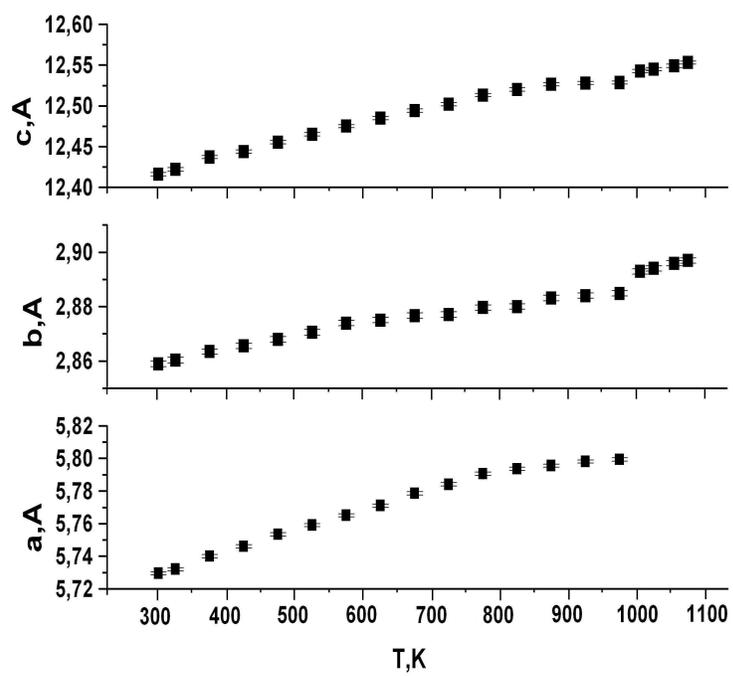

**Figure 5**

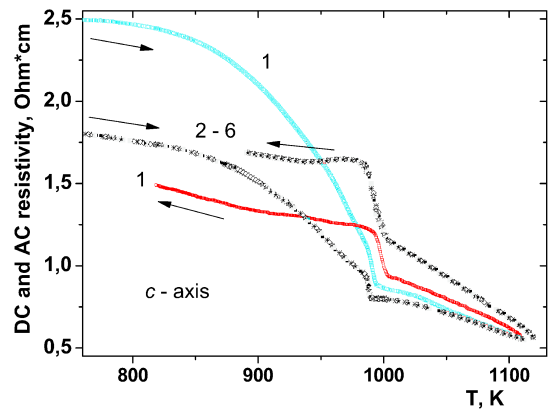

a)

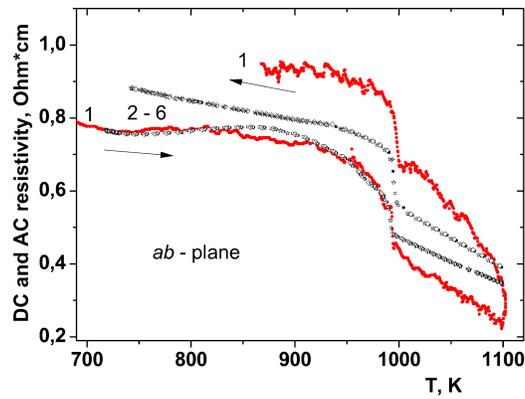

b)

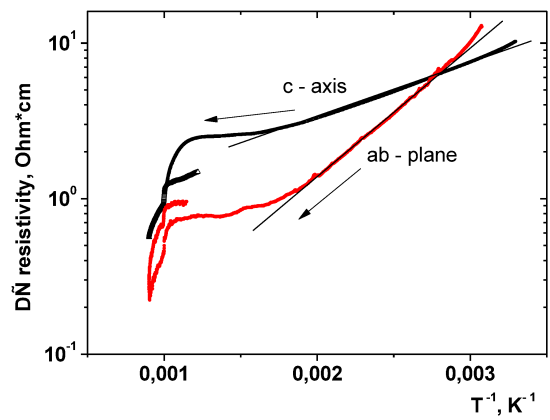

c)

**Figure 6**